\documentclass[10pt,aps,prd,superscriptaddress,nofootinbib,nobibnotes,longbibliography,floatfix,twocolumn]{revtex4-2}

\usepackage{bm}
\usepackage{mathtools,
amsmath,
amssymb,
amsfonts,
mathrsfs,
chngcntr,
multirow}

\let\cc\corresponds
\let\corresponds\relax
\usepackage{mathabx}
\let\corresponds\cc

\usepackage[utf8]{inputenc}
\usepackage[T1]{fontenc}
\usepackage[dvipsnames]{xcolor}
\usepackage{amsmath}
\usepackage[unicode]{hyperref}
\hypersetup{colorlinks=true, citecolor=MidnightBlue,
            linkcolor=MidnightBlue, urlcolor=MidnightBlue, linktocpage=true}
\usepackage[normalem]{ulem}
\usepackage{orcidlink}
\usepackage[capitalize]{cleveref}
\usepackage{float}
\usepackage{comment}
\usepackage{enumitem}

\usepackage{orcidlink}

\newcommand{\orcid}[1]{\href{https://orcid.org/#1}{\includegraphics[width=10pt]{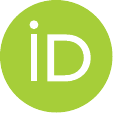}}}

\newcommand{\qm}[1]{``#1''}
\newcommand{\dd}{{\rm d}}

\begin{document}

\title{Discrete symmetries of modified Teukolsky equations}

\author{Ciro De Simone \orcid{0009-0004-0610-1686}}
\email{ciro.desimone@unina.it}
\affiliation{Dipartimento di Fisica \qm{E. Pancini}, Università di Napoli \qm{Federico II}, Complesso Universitario di Monte Sant' Angelo, Via Cinthia Edificio 6, I-80126 Napoli, Italy}
\affiliation{Istituto Nazionale di Fisica Nucleare, Sezione di Napoli, Complesso Universitario di Monte Sant' Angelo, Via Cinthia Edificio 6, 80126 Napoli, Italy}

\begin{abstract}

The Teukolsky equation possesses discrete symmetries that constrain the properties of black hole perturbations and their quasinormal mode spectrum. In this study, we explore how a class of modifications of the Teukolsky potential can alter the symmetry structure of the equation and break the $m=0$ degeneracy of quasinormal modes. We prove this result in the frequency domain using the master Teukolsky equation and in the time domain via $(2+1)$-dimensional simulations. We also show that the discrete symmetries can be leveraged for a more efficient characterization of the $m=0$ quasinormal modes from time-domain evolutions.  As a theory-specific application, we consider the case of higher-derivative theories of gravity, highlighting that time-domain implementations of frequency-domain potentials can give rise to additional nonphysical branches of modes.

\end{abstract}

\maketitle

\section{Introduction}

In general relativity (GR), the coalescence of two black holes (BHs) gives rise to a perturbed Kerr BH~\cite{PhysRevLett.11.237} which relaxes toward an equilibrium configuration via the emission of gravitational waves (GWs)~\cite{Maggiore:2007ulw,Maggiore:2018sht}. Since the first detection in 2015~\cite{LIGOScientific:2016aoc, LIGOScientific:2020ibl, KAGRA:2021vkt,LIGOScientific:2025slb}, GWs have paved the way to the exploration of gravity in the strong field regime, allowing precise tests of GR and deep insights into the nature of compact objects \cite{LIGOScientific:2025wao, LIGOScientific:2025rid}.

In the last phase of the coalescence, the so-called BH ringdown, GWs can be described in terms of quasinormal modes (QNMs)~\cite{Chandrasekhar:1985kt,Kokkotas:1999bd, Nollert:1999ji,Konoplya:2011qq, Franchini:2023eda}, which constitute the characteristic modes of vibration of a perturbed BH. In GR, the QNMs can be obtained from a master equation known as the Teukolsky equation~\cite{Teukolsky1972}, which governs BH perturbations. Moreover, the no-hair conjecture~\cite{Cardoso:2016ryw,Isi:2019aib} suggests that astrophysical BHs are fully characterized by three parameters: mass, charge and angular momentum. This observation is crucial in the so-called BH spectroscopy program~\cite{Detweiler:1980gk,Berti:2009kk, Berti:2025hly} which aims at characterizing BHs in terms of their QNMs. 

In order to probe deviations from GR and the Kerr QNM spectrum, a number of parametrized frameworks have been developed in the literature. Here, a deformation is introduced at the level of the Teukolsky equation~\cite{Cano:2024jkd, Tang:2025qaq, Yu:2025wpb} for rotating BHs or the Regge-Wheeler/Zerilli equation for the spherically symmetric case~\cite{Cardoso:2019mqo, McManus:2019ulj,Volkel:2022aca,Franchini:2022axs}. This approach is suitable to describe small deformations induced by beyond GR theories and is also motivated by recent works in effective field theories~\cite{Li:2022pcy,Hussain:2022ins,Cano:2023tmv}. 

For the purpose of this paper, it is important to point out that the Teukolsky equation exhibits different types of symmetries: continuous symmetries associated with time and azimuthal translations~\cite{Toth:2018qrx}, and discrete symmetries under combinations of complex conjugation, angular momentum, and coordinate reversals~\cite{Teukolsky1972, Krivan1997}. 

The aim of this paper is to discuss how the discrete symmetries are modified in frequency and time domains if a deformation is added to the Teukolsky equation. In particular, we identify a class of discrete transformations that preserve the modified Teukolsky equation and investigate the implications on the numerical time evolutions and QNM spectrum.

In order to check the validity of the results, we perform $(2+1)$-dimensional evolutions of the modified Teukolsky equation by simulating the scattering of a Gaussian wave packet with the deformed BH~\cite{Vishveshwara:1970zz, Krivan1997,Pazos-Avalos2004, Harms:2013ib}. This allows us to extensively test the discrete symmetries across many values of BH angular momentum, multipole number, and spin of the field. 

We find that the modification in the Teukolsky potential can break the degeneracy in the modes of azimuthal number $m=0$, leading to the excitation of two distinct sets of modes. The discrete symmetries are further tested for specific theories beyond GR such as higher-derivative theories of gravity~\cite{Cano:2024ezp}. This enables the investigation of a larger class of deformed potentials where the coefficients depend explicitly on the multipole numbers $(n,\ell,m)$.

Time-domain simulations have also been used to obtain QNM estimates for the modified Teukolsky equation via the Prony method~\cite{Berti2007}, obtaining remarkable agreement with the frequency-domain predictions. Here, we show that the discrete symmetries allow for a more efficient investigation of the $m=0$ fundamental modes.

The manuscript is organized as follows. In Sec.~\ref{sec: II}, we outline the general properties of the modified Teukolsky equation in frequency and time domains. In Sec.~\ref{sec: III}, we describe the discrete symmetry properties of the Teukolsky equation and its modified version, as well as the generalization of those results to a class of potentials that depend on the multipole $(n,\ell,m)$. In Sec.~\ref{sec: IV}, we test those symmetries via time-domain simulations and QNM computation, and study the case of higher-derivative theories of gravity. Finally, we draw our conclusions in Sec.~\ref{Conclusions}. In this paper, we adopt units in which $G_N=c=1$.

\section{Methods}
\label{sec: II}

This section outlines the modified Teukolsky framework in the frequency domain~(Sec.~\ref{sec: IIA}) and the implementation in the time domain~(Sec.~\ref{sec: IIB}).

\subsection{Frequency domain}
\label{sec: IIA}

In the parametrized quasinormal mode framework for modified Teukolsky~\cite{Cano:2024jkd}, the frequency-domain perturbation equation is expressed in Boyer-Lindquist coordinates as
    \begin{equation}\label{Mod_Teu_equation}
        \frac{1}{\Delta^s}\frac{d}{dr}\left[\Delta^{s+1}\frac{dR}{dr}\right]+[V(r)+\delta V(r)]R(r)=0\,.
    \end{equation}
Here, $R(r)$ is the radial component of a spin $s$ field and the radial Teukolsky potential is given by
\begin{equation}
    V(r) = 2is\frac{dK(r)}{dr}-\lambda_{\ell m}+\frac{1}{\Delta}\left(K(r)^2-isK(r)\frac{d\Delta}{dr}\right),
\end{equation}
while
\begin{align}
    \Delta &= r^2-2Mr+a^2,\\
    K(r)&=(r^2+a^2)\,\omega-am,\\
    \lambda_{\ell m} &= B_{\ell m}+a^2\omega^2-2am\omega,
\end{align}
are functions that depend on the mass $M$ and angular momentum $a$ of the BH, the frequency $\omega$, the azimuthal number $m$, and the separation constant $B_{\ell m}$.

In order to account for deviations from GR, a deformation $\delta V$ is introduced in the potential
    \begin{equation}\label{Mod_Teu}
        \delta V(r)=\frac{1}{\Delta}\sum_{k=-K}^{4}\alpha^{(k)}\left(\frac{r}{r_+}\right)^k,
    \end{equation}
which depends on the position of the outer event horizon of the Kerr BH $r_+ = M+\sqrt{M^2-a^2}$, a positive integer $K$~\cite{Kimura:2020mrh} and a set of dimensionful coefficients $\alpha^{(k)}$ that can be complex $\alpha^{(k)}=\alpha^{(k)}_R+i\alpha^{(k)}_I$. Those encapsulate deviations from GR  and $\alpha^{(k)}=0$ corresponds to the Teukolsky equation. This type of deformed potential is found in theories of gravity that assume small coupling corrections to GR~\cite{Cano:2024jkd,Cano:2023tmv}.

In Ref.~\cite{Cano:2024jkd}, the QNMs of the modified Teukolsky equation have been computed at first order in the complex coupling parameter $\alpha^{(k)}$ using Leaver's method~\cite{Leaver1985}
    \begin{equation}\label{omega_Mod}
         \omega_{n\ell m}\simeq \omega^0_{n\ell m}+\sum_{k=-K}^4 d^{(k)}_{\omega,n\ell m}\alpha^{(k)},
    \end{equation}
and they are identified by the overtone number $n$, the multipole indices $\ell$ and $m$. Moreover, $\omega^0_{n\ell m}$ corresponds to the GR QNMs and the deviation is expressed in terms of a set of coefficients $d^{(k)}_{\omega,n\ell m}$ which depend on the mode under consideration. The explicit expression of those coefficients can be found in a \texttt{GitHub} repository~\cite{github}.

\subsection{Time domain}
\label{sec: IIB}

In time domain, the general form of the vacuum Teukolsky equation for a field $\psi(t,r,\theta,\phi)$ is~\cite{Teukolsky1972, Krivan1997}:

    \begin{align}\label{master_equation_Kerr}
        &\left[\frac{(r^2+a^2)^2}{\Delta}-a^2\sin^2\theta\right]\frac{\partial^2\psi}{\partial t^2}+
        \left[\frac{a^2}{\Delta}-\frac{1}{\sin^2\theta}\right]\frac{\partial^2\psi}{\partial \phi^2}+ \nonumber\\& \frac{4Mar}{\Delta}\frac{\partial^2\psi}{\partial t\partial\phi}-2s\left[\frac{M(r^2-a^2)}{\Delta}-r-ia\cos\theta\right]\frac{\partial \psi}{\partial t}+
\nonumber\\&-\frac{1}{\sin\theta}\frac{\partial}{\partial \theta}\left(\sin\theta\frac{\partial\psi}{\partial\theta}\right)-2s\left[\frac{a(r-M)}{\Delta}+\frac{i\cos\theta}{\sin^2\theta}\right]\frac{\partial \psi}{\partial \phi}+\nonumber\\
        & -\Delta^{-s}\frac{\partial}{\partial r}\left(\Delta^{s+1}\frac{\partial \psi}{\partial r}\right)
        +(s^2\cot^2\theta-s)\psi=0,
    \end{align}
for scalar $(s=0)$, electromagnetic $(s=-1)$, and gravitational $(s=-2)$ perturbations. Moreover, Eq.~\eqref{master_equation_Kerr} is related to Eq.~\eqref{Mod_Teu_equation} via the ansatz $\psi=e^{-i\omega t} e^{im\phi}S(\theta)R(r)$ based on the separability of the Teukolsky equation~\cite{Teukolsky1972}.

Equation~\eqref{master_equation_Kerr} can also be recast in a compact form 
    \begin{equation}\label{Teu_compact}
        (\mathcal{O}+V_T)\,\psi(t,r,\theta,\phi)=0,
    \end{equation}
where the operator $\mathcal{O}$ collects all derivatives with respect to the variables $(t,r,\theta,\phi)$ and $V_T$ the terms that multiply the field $\psi(t,r,\theta,\phi)$,
    \begin{equation}
        V_T=s^2\cot^2\theta-s,
    \end{equation}
and characterizes perturbations of nonzero spin $s$. Taking into account the modification in the potential, the Teukolsky equation becomes
    \begin{equation}\label{mod_Teu_compact}
        (\mathcal{O}+V_T'(r))\,\psi(t,r,\theta,\phi)=0,
    \end{equation}
where 
    \begin{equation}
        V_T'(r)=V_T-\delta V(r).
    \end{equation}
It is interesting to notice that the deformation in Eq.~\eqref{Mod_Teu_equation} affects only the last term in Eq.~\eqref{master_equation_Kerr}, i.e., the term proportional to the field $\psi$. This is a consequence of the small coupling approximation, which allows one to separate radial and angular terms, as well as remove couplings among different multipoles $(\ell,m)$~\cite{Cano:2023tmv}.

There are several advantages in a time-domain formulation of the problem~\cite{Vishveshwara:1970zz}. First, simulations can probe the full linear perturbative response of the BH, from the early-time behavior associated with the propagation of initial data, the QNM ringing~\cite{DeSimone:2026mkz, Thomopoulos:2025nuf, Konoplya:2025afm} and the late-time tail~\cite{Price:2004mm, Rosato:2025rtr}. Moreover, they do not require the Teukolsky equation to be separable, even though the simulations become more involved.

In order to study the perturbative response of the modified Kerr BH, we evolve initial data corresponding to an ingoing Gaussian wave packet modulated by a spin-weighted spheroidal harmonic of multipole $(\ell,m)$. The time-domain integration is performed via the modified Lax-Wendroff method, which generalizes the code used in Ref.~\cite{DeSimone:2026mkz} for scalar perturbations~\cite{Doneva:2020nbb,Pedrotti:2024znu} and is second-order convergent~\cite{Krivan1997}. 
The details on the numerical implementation of the modified Teukolsky equation can be found in Ref.~\cite{DeSimone:2026waz}.

\begin{figure*}
\includegraphics[width= \linewidth]{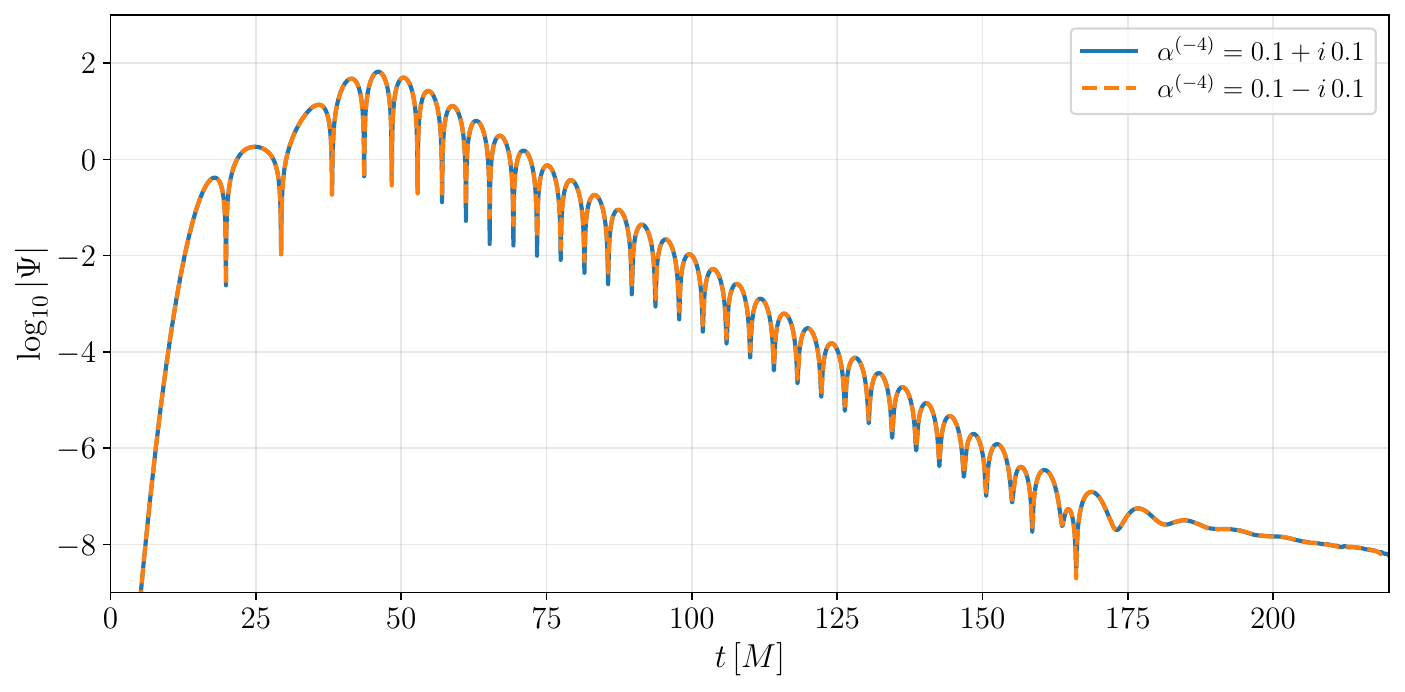}
\centering
\caption{Logarithmic plot of the field $\Psi$ for $a=0.25,\,s=-2,\,\ell=2,\,m=0,\,k=-4$ in units $M=0.5$ and complex conjugate values of $\alpha^{(-4)}=0.1+ i\,0.1$.}
\label{fig:m0_symmetries}
\end{figure*}

\section{Results}
\label{sec: III}

In this section, we first describe the symmetry properties of the Teukolsky equation in frequency and time domains (Sec.~\ref{subsec: IIIA}) and in Sec.~\ref{subsec: IIIB} how those symmetries are affected in the modified Teukolsky framework. Section~\ref{subsec: IIIC} discusses the discrete symmetries in the case of deformations that depend on the multipole $(n,\ell,m)$.

\subsection{Teukolsky equation}
\label{subsec: IIIA}

Let us first consider the Teukolsky equation in the frequency domain Eq.~\eqref{Mod_Teu_equation} with $\delta V(r)=0$. It is well known that, for a given $(\ell,\,m)$, the Teukolsky equation admits two sets of QNMs with different frequencies and damping times~\cite{Berti:2005ys, Dorband:2006gg,Ghosh:2024het}. The real and imaginary parts of those modes are related as follows
    \begin{align}
        \Re\,[\omega^{(i)}_{n\ell m}]&=-\Re\,[\omega^{(j)}_{n\ell-m}]\label{omegaR_sym},\\
        \Im\,[\omega^{(i)}_{n\ell m}]&=\Im\,[\omega^{(j)}_{n\ell-m}]\label{omegaI_sym},\\
        B^{(i)*}_{n\ell m}&=B^{(j)}_{n\ell-m} \label{sep_const_sym},
    \end{align}
where the indices $i,j\neq i$ denote the two sets of QNMs. They differ by the sign of the real part and can be identified with the multipoles $(\ell,m)$ and $(\ell,-m)$. In the eikonal picture of QNMs~\cite{Cardoso:2008bp, Yang:2012he}, these modes are associated with prograde photon orbits that are corotating $(m>0)$ or retrograde orbits counterrotating $(m<0)$ with the BH, respectively. Moreover, if $m=0$ the two modes have the same frequency (up to a minus sign) and damping time, thus prograde and retrograde modes become degenerate. 

This symmetry property can be obtained from the Teukolsky equation in the frequency domain. In fact, the equations for $(\ell,m)$ and $(\ell,-m)$ are connected by performing the following set of transformations~\cite{Krivan1997}:
\begin{enumerate}[label=(\roman*)]
    \item complex conjugation,
    \item $\omega \to -\omega^*$,
    \item $\theta \to \pi-\theta$.
\end{enumerate}
A consequence of those considerations is that time-domain simulations of an initial multipole $(\ell,m)$ will always excite the mode $(\ell,-m)$ and vice versa.

This result can be proved also starting directly from the time-domain Teukolsky equation equation~\eqref{Teu_compact}. In fact, the Teukolsky operator $(\mathcal{O}+V_T)$ is invariant under the combination of
\begin{enumerate}[label=(\roman*)]
    \item complex conjugation,
    \item $\phi\to -\phi$,
    \item $a \to -a$,
\end{enumerate}
which leads to the equation
    \begin{equation}
        (\mathcal{O}+V_T)\,\psi^*(t,r,\theta,-\phi)=0.
    \end{equation}
Thus, fields of opposite values of $a$ and reversed azimuthal coordinate $\phi$ can be identified with the multipole $(\ell,m)$ and $(\ell,-m)$,
    \begin{align}
        \psi_{\ell\, m}(t,r,\theta,\phi) &= \psi(t,r,\theta,\phi),\\
        \psi_{\ell\, -m}(t,r,\theta,\phi) &= \psi^*(t,r,\theta,-\phi).
    \end{align}

The results do not change significantly if one specifies the azimuthal dependence of the field $\psi=e^{-i\omega t} e^{im\phi}S(\theta)R(r)$, exploiting the axisymmetry of the spacetime, or chooses the tortoise and Kerr azimuthal coordinate $(r_*,\tilde\phi)$,
    \begin{equation}\label{Kerr_coord}
          \frac{\dd r_*}{\dd r}=\frac{r^2+a^2}{\Delta}\, \;\;\;\;\;\; \textrm{and} \;\;\;\;\;\; \dd\tilde\phi=\dd\phi+\frac{a}{\Delta}\dd r.
    \end{equation}
In fact, the symmetry transformations reduce to
\begin{enumerate}[label=(\roman*)]
    \item complex conjugation,
    \item $a \to -a$,
\end{enumerate}
while $\phi$ and $m$ are unaffected.

\subsection{Modified Teukolsky equation}
\label{subsec: IIIB}

Let us now investigate the effect of the deformation equation~\eqref{Mod_Teu} to the Teukolsky potential. 

From a frequency-domain perspective, if the coefficients $\alpha^{(k)} \in \mathbb{C}$, complex conjugation will couple the multipole $(\ell,m,\alpha^{(k)})$ to $(\ell,-m,\alpha^{(k)*})$. As a consequence,~\cref{omegaR_sym,omegaI_sym} become
    \begin{align}
        \Re\,[\omega^{(i)}_{n\ell m}(\alpha^{(k)})]&=-\Re\,[\omega^{(j)}_{n\ell-m}(\alpha^{(k)*})], \label{omegaR_sym_Mod}\\
        \Im\,[\omega^{(i)}_{n\ell m}(\alpha^{(k)})]&=\Im\,[\omega^{(j)}_{n\ell-m}(\alpha^{(k)*})].
    \end{align}
In order to study the effect on time-domain evolutions, let us distinguish two possibilities in Eq.~\eqref{mod_Teu_compact}: $\alpha^{(k)} \in \mathbb{R}$ and $\alpha^{(k)}\in\mathbb{C}$. In the first case, the modified potential $V_T'$ is real, thus the symmetry transformation described above holds
    \begin{equation}
        (\mathcal{O}+V_T')\,\psi^*(t,r,\theta,-\phi)=0,
    \end{equation}
and the two modes evolved in the time domain correspond to the same value of $\alpha^{(k)}$: $(\ell,m,\alpha^{(k)})$ and $(\ell,-m,\alpha^{(k)})$. Therefore, for $m=0$ the modes are degenerate since they correspond to the same deformation. 

On the other hand, if $\alpha^{(k)}_I\neq0$, the potential becomes complex and it is no longer invariant under the discrete symmetries of the Teukolsky equation
    \begin{equation}
        V_T'^*\neq V_T', 
    \end{equation}
thus $\psi(t,r,\theta,\phi)$ and $\psi^*(t,r,\theta,-\phi)$ do not satisfy the same differential equation. Nonetheless, the modified Teukolsky potential is still invariant if one performs the additional transformation 
\begin{enumerate}[label=(\roman*)]\label{new_sym}
    \item $\alpha^{(k)} \to \alpha^{(k)\,*}$,
\end{enumerate}
which is equivalent to changing the sign of the imaginary part of $\alpha^{(k)}$, i.e., $\alpha^{(k)}_I\to - \alpha^{(k)}_I$. As a consequence, the two modes evolved in the numerical simulations correspond to the pairs $(\ell,m,\alpha^{(k)})$ and $(\ell,-m,\alpha^{(k)*})$. It is also worth stressing that those results are valid independent from the spin $s$ of the perturbed field and the index $k$.

This implies that the $m=0$ modes are no longer degenerate since they correspond to $\alpha^{(k)}$ and its complex conjugate $\alpha^{(k)*}$. These modes will, in general, have different real and imaginary part.

The symmetries of the time-domain Teukolsky equation are directly reproduced in the numerical simulations. For a given $\ell$, the $m=0$ waveforms are well described as combinations of the fundamental mode $(n=0)$ associated with $\alpha^{(k)}$ and its complex conjugate $\alpha^{(k)*}$. 

Figure~\ref{fig:m0_symmetries} displays two waveforms corresponding to the multipole $\ell=2,\, m=0$, and $\alpha^{(-4)}=0.1\pm i\,0.1$. As expected, the two profiles coincide since the same pair of modes is excited in each simulation, with similar amplitudes and phases.

The frequencies extracted using the Prony method~\cite{Berti2007}, which consists of fitting the numerical waveform with a sum of damped sinusoids, agree to less than $0.05\%$. Moreover, the Prony estimates can be compared with the frequency-domain QNM predictions given by Eq.~\eqref{omega_Mod} and, for those values of $\alpha^{(-4)}$, the relative error for the fundamental mode is $\leq1\%$. 

The symmetries of the modified Teukolsky equation have been further tested for several values of $a,\,\ell,\,k$, and $s=0,-1,-2$. In all those cases, the simulations are in agreement with the theoretical predictions and show the qualitative behavior of Fig.~\ref{fig:m0_symmetries}.

\subsection{Multipole-dependent deformations}
\label{subsec: IIIC}
In the previous subsections, the coefficients $\alpha^{(k)}$ have been treated as pure complex numbers. It is interesting to notice, however, that in some theory-specific applications (e.g., higher-derivative theories of gravity~\cite{Cano:2024ezp}) the deformation coefficients will, in general, depend also on the multipole $(n,\ell,m)$, i.e., $\alpha^{(k)}=\alpha_{n\ell m}^{(k)}$. As a consequence, they will also be affected by the symmetry transformations. 

From a physical perspective, this implies that the potential deformation has a different form $\delta V(\alpha^{(k)}_{n\ell m})$ for prograde and retrograde modes, resulting in different Teukolsky equations for each multipole. Here, the notation $\delta V(\alpha^{(k)}_{n\ell m})$ means that the deformed potential depends on a combination of $\alpha^{(k)}_{n\ell m}$ coefficients. From a mathematical standpoint, instead, it is crucial to observe that any deformation of the potential defines a new Teukolsky operator, with its own QNM spectrum and perturbative properties. 

This implies that, in numerical time evolutions, the potentials for the multipoles $(n,\ell,m)$ and $(n,\ell,-m)$ will excite, respectively
    \begin{align}
        \delta V(\alpha^{(k)}_{n\ell m})&: \omega_{n\ell m}(\alpha^{(k)}_{n\ell m}),\,\omega_{n\ell -m}(\alpha^{(k)*}_{n\ell m}),\label{deltaV_pro}\\
        \delta V(\alpha^{(k)}_{n\ell -m})&: \omega_{n\ell -m}(\alpha^{(k)}_{n\ell -m}),\,\omega_{n\ell m}(\alpha^{(k)*}_{n\ell -m}).\label{deltaV_retro}
    \end{align}
Those results can be summarized as follows: any prograde ($m>0$) deformation of the potential will excite the corresponding prograde mode $\omega_{n\ell m}(\alpha^{(k)}_{n\ell m})$, as well as the retrograde mode of the prograde potential $\omega_{n\ell -m}(\alpha^{(k)*}_{n\ell m})$, and vice versa for retrograde ($m<0$) deformations. From the conditions~\eqref{deltaV_pro} and \eqref{deltaV_retro} and in the general case $\alpha_{n\ell m}^{(k)}\neq\alpha_{n\ell -m}^{(k)}$, the criterion that the coefficients must satisfy for the two pairs of modes to coincide is
    \begin{equation}\label{cond_deg}
        \alpha_{n\ell m}^{(k)}=\alpha_{n\ell -m}^{(k)*}.
    \end{equation}
This guarantees that the modes $\omega_{n\ell -m}(\alpha^{(k)*}_{n\ell m})$ excited by the potential $\delta V(\alpha^{(k)}_{n\ell m})$ correspond to the  mode $\omega_{n\ell -m}(\alpha^{(k)}_{n\ell -m})$ associated with the potential $\delta V(\alpha^{(k)}_{n\ell -m})$. 

Therefore, Eq.~\eqref{cond_deg} can, in principle, be used to identify specific classes of deformed potentials. If it is not satisfied, the additional modes excited in the time-domain simulation will be spurious modes that contaminate the waveform. They can in principle even be unstable, depending on the value of the deformation coefficient and should thus be taken into account when studying the numerical evolutions. Clearly, those instabilities are nonphysical.

It is also important to stress that, if $m=0$, prograde and retrograde potentials coincide and there is no distinction between the sets of QNMs excited in the numerical evolutions. The same argument implies that, if $\alpha_{n\ell m}^{(k)}=\alpha_{n\ell -m}^{(k)}$, simulations for opposite values of $m$ will coincide only for real values of $\alpha_{n\ell m}^{(k)}$. Otherwise, modes of opposite $m$ will be associated with complex conjugate values of $\alpha^{(k)}_{n\ell m}$. 

\section{Applications}
\label{sec: IV}

In this section, we discuss applications of the discrete symmetries to the computation of the $m=0$ QNMs~(Sec.~\ref{Sec. IVA}) and to the deformed Teukolsky potentials found in higher-derivative theories of gravity~(Sec.~\ref{Sec. IVB}).
\subsection{QNM computation}
\label{Sec. IVA}

As an application of the discrete symmetries discussed in the previous sections, let us focus on the $n=0, m=0$ modes of the modified Teukolsky equation in the case where the $\alpha^{(k)}$ coefficients are complex but independent of $(n,\ell,m)$. 

In the frequency domain, the QNMs can be computed using Leaver's method~\cite{Leaver1985} or via Eq.~\eqref{omega_Mod} at linear order in  $\alpha^{(k)}$~\cite{Cano:2024jkd}; whereas in the time domain they can be directly extracted from the waveforms using the Prony method. More importantly, the numerical simulations can be used to obtain an independent estimate of the coefficients $d^{k}_{\omega,0\ell0}$ in Eq.~\eqref{omega_Mod} or the higher-order contributions in $\alpha^{(k)}$. 

This can be achieved by the following procedure: (i) select one value of $k$ in the deformed potential given by Eq.~\eqref{Mod_Teu}, (ii) extract the QNMs of multiple simulations for values of $\alpha^{(k)}$ around zero and (iii) fit the real and imaginary parts of the QNMs with a polynomial in $\alpha^{(k)}$. 

In order to determine the linear coefficients, one can select multiple equally spaced real values of $\alpha^{(k)} = \alpha^{(k)}_R \in [-0.1,0.1]$ and perform numerical evolutions for each of those values. Since the deformation parameters are real, each simulation will provide information on only one mode corresponding to $\alpha^{(k)}_R$.

On the other hand, if one takes purely imaginary values of $\alpha^{(k)} = i\,\alpha^{(k)}_I$, the discrete symmetries of the modified Teukolsky equation allow one to select only $\alpha^{(k)}_I \geq 0$ (or $\alpha^{(k)}_I \leq 0$) since all the complex conjugate values are already included in the time-domain simulations. The symmetries thus reduce the number of simulations by a factor of $2$.

\begin{figure}[h]
\includegraphics[width= 0.8\linewidth]{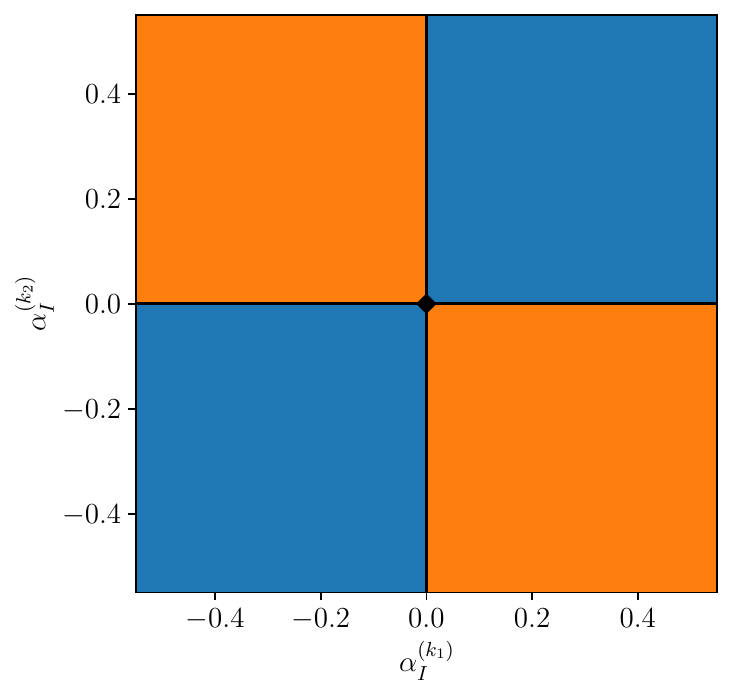}
\centering
\caption{Quadrants of the same color are mapped by the symmetry transformation for $m=0$. Along the main axes $\alpha_I^{(k_1)}=0$ and $\alpha_I^{(k_2)}=0$ identified by the black lines, opposite values of $\alpha_I^{(k_1)}$ (or $\alpha_I^{(k_2)}$) are also mapped by the symmetry.}
\label{fig:symmetry_map}
\end{figure}

It is important to highlight that the Prony method does not \textit{a priori} specify which of the two extracted frequencies correspond to $\alpha^{(k)}_I$ and to $\alpha^{(k)*}_I$. In order to separate them, one has to observe that, according to Eq.~\eqref{omegaR_sym_Mod}, the two modes have frequencies of opposite sign and can be identified directly with opposite values of $\alpha^{(k)}_I$: the positive frequency being associated with $\alpha^{(k)}_I$ and the negative frequency with $\alpha^{(k)*}_I$. In this regard, the $m=0$ modes of the modified Teukolsky equation for complex $\alpha^{(k)}$ behave exactly like the $m\neq0$ modes of the Kerr BH.

This procedure can be generalized to the case where multiple $k$ are nonzero in Eq.~\eqref{Mod_Teu}. For example, let us assume that there are two nonzero contributions in Eq.~\eqref{Mod_Teu}, labeled as $k_1$ and $k_2$. Based on the previous considerations, the symmetries map all the values of $\alpha_I^{(k_1)}$ and $\alpha_I^{(k_2)}$ with their complex conjugates. In terms of the plane $(\alpha_I^{(k_1)},\,\alpha_I^{(k_2)})$ shown in Fig.~\ref{fig:symmetry_map}, the symmetries connect all the points that lie on even or odd quadrants, reducing drastically the number of simulations required to estimate the coefficients, especially the higher-order ones. This result can be further generalized if more than two nonzero values of $k$ are considered in the modified Teukolsky potential. 

\subsection{Higher-derivative theories of gravity}
\label{Sec. IVB}

An example of the multipole-dependent deformation presented in Sec.~\ref{subsec: IIIC} can be found in higher-derivative theories of gravity (HDG)~\cite{Cano:2024ezp,Maenaut:2024oci}. Here, the first-order modified Teukolsky potential can be expressed as in Eq.~\eqref{Mod_Teu} with four non-zero complex coefficients $\alpha^{(k)}$, corresponding to $k=-2,0,1,2$. Those coefficients depend on $(n,\ell,m)$ because they are functions of the Kerr QNM frequencies. HDG thus constitutes an ideal testbed for the discrete symmetries of modified Teukolsky equations.

This test has been realized by implementing the first order HDG modified Teukolsky equation in the time domain using the formalism introduced in~\cite{DeSimone:2026waz} and evaluating the coefficients at a fixed Kerr QNM. The explicit expression of the modified potential can be found in a~\texttt{GitHub} repository~\cite{github2} for many values of $(\ell,m)$, different orders in the angular momentum expansion, and different types of HDG modifications of GR. Moreover, the QNMs extracted from the waveforms show good agreement with the theoretical predictions~\cite{Cano:2024jkd,Cano:2024ezp}. For small values of $m$, the relative errors on both the real and imaginary parts of the fundamental mode are of order of a few percent~\cite{DeSimone:2026waz}.

For the $m=0$ modes, since the deformation coefficients are complex, the time-domain evolutions exhibit the same degeneracy breaking described in Sec.~\ref{subsec: IIIB}. It is important to stress, however, that the modes associated with the complex conjugate deformations $\alpha^{(k)*}$ are physically relevant only if the coefficients $\alpha^{(k)*}$ belong to the parameter space of HDG at first order in the coupling parameter.

In the case $m\neq0$, the condition~\eqref{cond_deg} requires the knowledge of the modified potentials corresponding to retrograde modes. Those can be directly obtained from Ref.~\cite{github2} considering that, as in GR, the retrograde modes correspond to negative values of angular momentum~\cite{Cano:2024ezp}. Following this strategy, it is found that the condition~\eqref{cond_deg} is, in general, not satisfied in HDG at first order in the coupling parameter. This implies that the simulations for given initial data $(n=0,\,\ell,m)$ do not provide physical information about both prograde and retrograde fundamental modes. In particular, the conditions~\eqref{deltaV_pro} and \eqref{deltaV_retro} indicate that the fundamental target mode $\omega_{0\ell m}$ can be mapped to the parametrized frequency-domain approach, while the conjugate mode $\omega_{0\ell -m}$ will be a spurious component. The time-domain simulations will also contain additional contributions, such as overtones and couplings between different multipoles. Those modes either decay very quickly or are weakly excited; thus, they cannot be reliably extracted from the waveforms. However, since they are associated with different frequencies in the Kerr limit, in general, they also constitute a contamination of the waveform and do not provide physical information.

\section{Conclusions}
\label{Conclusions}

Theories of gravity beyond GR~\cite{Capozziello:2011et, Nojiri:2017ncd} modify the gravitational field equations as well as the Teukolsky equation~\cite{Li:2022pcy, Cano:2024bhh, Cano:2023tmv}. In this paper, we have investigated how a deformation of the Teukolsky potential given by Eq.~\eqref{Mod_Teu} can impact the discrete symmetries of the Teukolsky equation. In addition, we have studied the effect of those modified symmetries on the time-domain evolution of the perturbation equation. 

First, we have implemented the parametrized framework for modified Teukolsky~\cite{Cano:2024jkd} in the time domain. This has allowed us to perform $(2+1)$-dimensional scattering experiments of Gaussian initial data with the modified BH and extract QNMs from the numerical waveforms. Then, we have identified a set of discrete symmetry transformations that leaves the Teukolsky equation invariant and generalized those results to the modified Teukolsky equation.  We have further discussed the effect of the discrete symmetries on the numerical evolutions for a specific deformed potential that depend on the multipole $(n,\ell,m)$.

By combining time- and frequency-domain techniques, we have shown that, unlike the GR case, the $m=0$ degeneracy between prograde and retrograde modes can be broken if the deformation is complex valued. In particular, $m=0$ simulations of modified Teukolsky equations can display a similar splitting as the $m\neq0$ modes of the Teukolsky equation. The discrete symmetries have also been tested via numerical simulations for several values of the BH angular momentum, spin of the field and angular multipole.  

As applications of these results, we have first presented a strategy to compute the $m=0$ fundamental QNMs of the modified Teukolsky equation which benefits from the discrete symmetries. Then, we have studied modified Teukolsky potentials found in higher-derivative theories of gravity. We have stressed that the additional modes excited in the simulations due to the discrete symmetries can be nonphysical and contaminate the waveforms. This shows that multipole-dependent potentials in the frequency domain cannot be automatically promoted to time-domain equations without creating branches whose physical status must be checked. 

It is important to highlight that the discrete symmetries analyzed in this paper hold for the parametrized framework for the modified Teukolsky equation~\cite{Cano:2024jkd}. However, the procedure can be readily generalized to other types of deformations, which might even be frequency dependent or nonseparable~\cite{Ghosh:2023etd}. 

In specific theories of gravity beyond GR, the deformation coefficients can also be more complicated functions of the indices $(\ell,\,m)$, for instance, if there is a coupling among different multipoles, which would further affect the discrete symmetries of the Teukolsky equation. Those models can, in principle, be implemented in time-domain simulations and provide a powerful and flexible way of investigating beyond GR effects on BH perturbations and QNMs.

\acknowledgments
C.D.S. would like to thank Sebastian H. V\"olkel, Nicola Franchini, and Vittorio De Falco for valuable discussions and feedback on the manuscript. C.D.S. is grateful to Kostas D. Kokkotas and the University of T\"ubingen for hospitality during the realization of this work. C.D.S. acknowledges the support of INFN, {\it sez. di Napoli}, {\it iniziativa specifica}  QGSKY. 

\bibliography{literature}

\appendix

\end{document}